\documentclass[10pt,a4paper,aps,prb,reprint,nobibnotes,floatfix,superscriptaddress,showpacs]{revtex4-1}

\usepackage[utf8]{inputenc}
\usepackage[english]{babel}

\usepackage{amsfonts}
\usepackage{amsmath}
\usepackage{amssymb}
\usepackage{bbold}
\usepackage[dvipsnames,svgnames,table]{xcolor}
\usepackage{graphicx}
\usepackage{fancyhdr}
\usepackage{float}
\usepackage{braket}
\usepackage{float}
\usepackage[caption=false]{subfig}
\usepackage{tikz}  
\usepackage{chngcntr}

\usepackage[dvipsnames]{xcolor}

\usepackage{hyperref}
\hypersetup{
    pdfstartview={FitH},
    colorlinks=true,    
    linkcolor=NavyBlue, 
    citecolor=Maroon,   
    filecolor=NavyBlue, 
    urlcolor=NavyBlue   
}

\makeatletter
\def\footnoterule{\kern -10pt
    \hrule \@width 100pt \kern 10pt} 
\makeatother

\begin{document}

\title{Domain wall damped harmonic oscillations induced by curvature gradients in elliptical magnetic nanowires}
\author{G. H. R. Bittencourt}
\affiliation{Departamento de F\'isica, Universidade Federal de Vi\c cosa, 36570-900, Vi\c cosa, Brazil}
\author{S. Castillo-Sep\'ulveda}
\affiliation{Departamento de Ingenier\'ia, Universidad Aut\'onoma de Chile, Avda. Pedro de Valdivia 425, Providencia, Chile}
\author{O. Chubykalo‐Fesenko}
\affiliation{Instituto de Ciencia de Materiales de Madrid, CSIC, Cantoblanco, 28049 Madrid, Spain}
\author{R. Moreno}
\affiliation{Earth and Planetary Science, School of Geosciences, University of Edinburgh, Edinburgh EH9 3FE, UK}
\author{D. Altbir}
\affiliation{Universidad de Santiago de Chile, Departamento de F\'isica, CEDENNA,      9170124, Santiago, Chile}
\author{V. L. Carvalho-Santos}
\affiliation{Departamento de F\'isica, Universidade Federal de Vi\c cosa, 36570-900, Vi\c cosa, Brazil}

\begin{abstract}
Understanding the domain wall (DW) dynamics in magnetic nanowires (NW) is crucial for spintronic-based applications demanding the use of DWs as information carriers. This work focuses on the dynamics of a DW displacing along a bent NW with an elliptical shape under the action of spin-polarized electric currents and external magnetic fields. Our results evidence that a curvature gradient  induces an exchange-driven effective tangential field responsible for pinning a DW near the maximum curvature point in a NW. The DW equilibrium position depends on the competition between the torques produced by the external stimuli and the curvature-induced effective fields.  When the external stimuli are below a certain threshold, the DW follows a damped harmonic oscillation around the equilibrium position. Above this threshold,  DW displaces along the NW under an oscillatory translational motion.
\end{abstract}
\pagebreak
        
\maketitle

\section{Introduction}

Magnetic nanowires (NW) are promising candidates to create a new generation of spin-based devices at the nanoscale. Their geometry-induced strong uniaxial anisotropy parallel to the NW axis \cite{Slatiskov} allows the nucleation of domain walls (DW) along the longer dimension that can be guided them through external magnetic fields and/or electric currents \cite{Berger,Wartele,Schobitz}. The proper control of the DW dynamics is a key point for developing technologies that use these magnetization textures as information carriers \cite{Racetrack,RAMemory}, nano oscillators \cite{Nanoosc-1,Nanoosc-2}, logic gates \cite{LGates,LogDev-1,LogDev-2}, and nanoantennas \cite{Espejo}. Consequently, controlling the DW dynamical response to external stimuli is a very hot topic in material sciences.

Tuning DW dynamical properties by means of geometrical considerations is a well-established strategy. One example in this direction is the shaping of the NW cross section. In magnetic nanostripes there exists a threshold for the intensity of the  external stimuli, known as Walker breakdown \cite{Walker,Mougin-DW},  at which the DW drastically alters its dynamical behavior. Below this limit, the DW propagates constantly  and its velocity is linearly proportional to the intensity of the stimuli. Above this limit - called Walker breakdown-, the DW  exhibits an oscillatory behaviour along the stimuli direction and the average velocity losses its linearity. By shaping the NW cross-section, the Walker limit can be either suppressed (circular cross-section \cite{Yan-PRL}) or engineered,  {\it i.e.} tunning the amplitude and frequency of the oscillations (polygonal cross sections \cite{Altbir-SciRep}).

Bending NWs is an alternative strategy to tune DW dynamical properties that has recently attracted lots of attention. In this case, the Walker regime is presented even for circular cross sections. In bent NWs the origin of the Walker breakdown is on the curvature-induced exchange torque, while in straight ones the dipolar energy coming from the cross section shape is the source of this effect. Importantly, the threshold for the Walker breakdown as well as the amplitude of the oscillations can be controlled by means of the curvature \cite{Moreno,Cacilhas,Yershov-2016}, but not the frequency. Nevertheless, combining both bending and shaping cross section, {\it e.g.}, bent nanostripes, can be used to control the frequency of the oscillations. An interesting result is that in bent nanostripes the external stimuli exhibit an extra threshold  in which the DW phase reorients \cite{Bittencourt-2021}.  

Introducing a curvature gradient in bent NWs provides extra and intriguing dynamical properties to the DW motion. In this case, it was shown that magnetization collective modes displacing along a low dimensional nanomagnet with a variable mean curvature suffers the effects of an effective force \cite{CS-APL} due to a curvature-induced effective anisotropy and  Dzyaloshinskii–Moriya interactions \cite{Gaididei-PRL,Makarov-PRL}. This curvature-induced effective force is responsible for exciting phenomena such as the attraction and/or repulsion of skyrmions by a curved bump in a nanoshell \cite{CS-APL,Korniienko,Kravchuk-2018} and the displacement of transverse DW along a helicoidal NW even in the absence of external fields or electric currents \cite{Yershov-2018,Skoric-2021}. In particular,  Yershov \textit{et al.} \cite{Yershov-2015} analyzed the existence of a pinning potential at the NW maximum curvature region in a torsionless NW with variable curvature. In that work, the authors studied the DW motion near the maximum curvature region, obtaining DW harmonic decaying oscillations \cite{Yershov-2015}. The works cited above evidence that an adequate choice of the NW geometric and magnetic parameters allows the control of the DW position, velocity, and phase.


 Motivated by these results, in this work we go deeper into the dynamics of a DW  displacing along an elliptically curved NW under the action of electric currents and magnetic fields. Since this system presents a curvature gradient, the emergence of an effective exchange-driven curvature-induced tangent magnetic field is responsible for pinning the DW near the maximum curvature region of the NW, as also observed by Yershov \textit{et al.} \cite{Yershov-2015}. We also show that the DW equilibrium position depends on the competition between the torques produced by the external stimuli and the exchange-driven curvature-induced tangential effective field. Additionally, we obtain that the DW follows a damped harmonic oscillation around an equilibrium position if the external stimuli are below a critical threshold. Above this threshold value, the DW propagates following an oscillatory motion along the NW with variable amplitude.

This work is organized as follows: Section \ref{Theory} presents the theoretical model adopted in this work. In section \ref{Result-1}, we present the DW dynamics in the absence of external stimuli while sections \ref{Result-2} and \ref{Result-3} describes the DW dynamics under an external current or magnetic field, respectively. Our conclusions are presented in section \ref{Conclusions}.

\section{Theoretical model}\label{Theory}

\subsection{General formulation}

The geometrical description of an arbitrary curvilinear NW without torsion can be done by adopting an orthogonal basis ($\mu, \eta, z$), where $\hat{\eta}$ is an azimuthal-like direction, which points tangent to the NW axis, and $\hat{\mu}$ is a radial-like unitary vector pointing outward the bent, perpendicular to the NW axis (see Fig.\ref{fig1}-a). The third unitary vector of the considered basis points along the $z$-axis direction, and can be defined from $\hat{z} = \hat{\mu}\times\hat{\eta}$ and therefore, vectors $\hat{\mu}$ and $\hat{\eta}$ lie in the $xy$ Cartesian plane. Under this framework, a NW can be described as $\boldsymbol{r}(\mu,\eta, z) = (x, y, z)$, where $x \equiv x(\mu,\eta)$ and $y \equiv y(\mu,\eta)$. In addition to the unitary vectors, it is important to determine the length element in a curvilinear orthogonal basis along an arbitrary direction $\hat{\xi}$, which is formally defined as $dq_{\xi} = h_{\xi}d\xi$. Here, 
$h_{\xi} = \sqrt{(\partial\mathbf{r}/\partial \xi)\cdot(\partial\mathbf{r}/\partial \xi)}$ is the modulus of the metric factor associated to the $\xi$-direction and $\hat{\xi} =  h^{-1}_{\xi}\partial\mathbf{r}/\partial \xi$.

In our study we consider a transverse DW with fixed shape and width during its displacement along the NW. Therefore, each magnetic moment of the system follows the same dynamics of the DW center. The transverse DW magnetization profile can be written in terms of the arbitrary curvilinear basis  as $\mathbf{m}=\sin\Omega\sin\phi\,\hat{\mu}+\cos\Omega\,\hat{\eta}+\sin\Omega\cos\phi\,\hat{z}$, where $\mathbf{m}=\mathbf{M}/M_s$, and $M_s$ is the saturation magnetization. The parameter $\Omega$ is the angle between $\mathbf{m}$ and the azimuthal-like unitary vector $\hat{\eta}$, while $\phi$ is the angle between the magnetization projection in the $\mu\, z$-plane and the vector $\hat{z}$, as shown in Fig.\ref{fig1}-a. Furthermore, as the dimensions of the NW cross-section are much smaller than its length, we can assume that the magnetization distribution varies only along the NW length ($\mathbf{m} = \mathbf{m}(\eta)$), and it is uniform along the $\hat{\mu}$ and $\hat{z}$ directions. This rigid transverse DW profile can be fitted by the ansatz $\Omega \equiv \Omega(\eta)=2\arctan\left\{\exp[(q_{\eta}-q_0)/\delta]\right\}$, where $q_{\eta}$ is an arbitrary position on the NW, $q_{0}$ defines the position of the DW center, and $\pi\delta$ is the DW width. It is worth noticing that the DW center is characterized by $\Omega = \pi/2$. From the adopted model, the DW velocity is determined by 

\begin{equation}\label{DW-Vel}
    v=\frac{dq_{0}}{dt} = h_{\eta}\frac{d\eta_{0}}{dt}=-\delta\left(\frac{d\Omega}{dt}\right)_{\Omega=\pi/2}\,.
\end{equation} 

The magnetization dynamics under an external magnetic field and/or an electric current is described by the Landau-Lifshitz-Gilbert (LLG) equation with spin transfer torque adiabatic and non-adiabatic terms 

\begin{eqnarray}
    \frac{d\mathbf{M}}{dt} = -\gamma\mathbf{M}\times\mathbf{H_{\text{eff}}} + \dfrac{\alpha}{M_s}\mathbf{M}\times\frac{d\mathbf{M}}{dt}-u\frac{\partial \mathbf{M}}{\partial q_{\eta}}\nonumber\\+\frac{\beta u}{M_s}\mathbf{M}\times\frac{\partial \mathbf{M}}{\partial q_{\eta}}\,,
    \label{LLG}
\end{eqnarray}

\noindent where $\gamma$ is the gyromagnetic ratio and  $\alpha$ is the damping parameter. Each term of the above equation is associated with the torque produced by effective magnetic fields or currents on the magnetization. The first and second terms in the right side of Eq. \eqref{LLG} are related to the torques produced by the effective field $\mathbf{H_{eff}}$ and the Gilbert damping, respectively. Here, the effective field can be written in a curvilinear basis as $\mathbf{H_{eff}} = H_{\text{eff} \, \mu}\,\hat{\mu} + H_{\text{eff} \, \eta}\,\hat{\eta} + H_{\text{eff} \, z}\,\hat{z}$. For simplicity we considered a Permalloy NW, therefore anisotropy vanishes and then, $\mathbf{H_{eff}}$ includes the external field $\mathbf{H}$, and the effective fields produced by exchange $\mathbf{H_{ex}}$ and dipolar $\mathbf{H_d}$ interactions, that will be presented after in this text. The third and fourth terms in the right side of Eq. \eqref{LLG} are associated respectively with the adiabatic and non-adiabatic spin transfer torques. Here, $\beta$ is the phenomenological non-adiabatic spin-transfer parameter, $u=gJ_e\mu_\text{B}P/2eM_s$ has velocity dimension and depends on an electric current $J_e$ injected into the magnetic sample, $g$ is the Land\'e factor, $\mu_{\text{B}}$ is the Bohr magneton, $P$ is the spin-polarization factor of the electric current, and $e$ is the electron charge. The adopted magnetic parameters for Permalloy  are \cite{Thiaville, Cacilhas, Bittencourt-2021} a saturation magnetization $M_s = 813 $ erg G$^{-1}$ cm$^{-3}$, an exchange stiffness $A=1.07 \times 10^{-6}$ erg/cm.  We use a damping constant $\alpha = 0.01$, and the non-adiabatic spin transfer torque parameter $\beta = 0.04$. Finally, we consider $\pi\delta = 33$ nm, which corresponds to the width of a transverse DW lying in a NW with a diameter of 30 nm \cite{Moreno-JMMM}.  

In the rigid transverse DW model, all terms in the LLG-equation can be compacted into the form $d\mathbf{M}/dt = -\gamma \boldsymbol{\Gamma}$, with $\boldsymbol{\Gamma}$ being the total torque resulting from all relevant interactions in the DW center. When written in terms of the ($\mu$, $\eta$, $z$)-coordinate system, the total torque is given by

	\begin{eqnarray}
     \boldsymbol{\Gamma}_{\mu,  \eta,  z} = M_s\begin{bmatrix} -\left(H_{\text{eff} \, \eta} +  \frac{\alpha}{\gamma}\dot{\Omega} + \frac{\beta u}{\gamma \delta} \right)\cos\phi \\  H_{\text{eff} \, \mu}  \cos\phi - H_{\text{eff} \, z}\sin\phi - \frac{\alpha}{\gamma}\dot{\phi} - \frac{u}{\gamma \delta} \\ \left(H_{\text{eff} \, \eta} +  \frac{\alpha}{\gamma}\dot{\Omega} + \frac{\beta u}{\gamma \delta} \right)\sin\phi
	\end{bmatrix}.
\end{eqnarray}

It is convenient to represent the total torque in terms of a spherical system ($\rho$, $\Omega$, $\phi$) lying in the curvilinear basis, as depicted in Fig. \ref{fig1}-a. For this purpose, the total torque in the DW center can be transformed as  $\boldsymbol{\Gamma} \equiv \boldsymbol{\Gamma}_{\rho,  \Omega,  \phi} = \mathcal{R}(\pi/2, \phi) \boldsymbol{\Gamma}_{\mu,  \eta,  z}$, where the transformation matrix $\mathcal{R}(\Omega, \phi)$ reads 

	$$\mathcal{R}(\Omega,\phi)=\begin{bmatrix}
		\sin\Omega\sin\phi & \cos\Omega & \sin\Omega\cos\phi  \\
		\cos\Omega\sin\phi & -\sin\Omega & \cos\Omega\cos\phi  \\
		\cos\phi & 0 & -\sin\phi
	\end{bmatrix} \, .$$
	
\noindent Therefore, the total torque can be rewritten as  

	\begin{eqnarray}\label{Torques}
     \boldsymbol{\Gamma}_{\rho,  \Omega,  \phi} = M_s\begin{bmatrix} 0 \\  H_{\text{eff} \, z}\sin\phi - H_{\text{eff} \, \mu}\cos\phi + \frac{\alpha}{\gamma}\dot{\phi} + \frac{u}{\gamma \delta} \\ - \left(\frac{\alpha}{\gamma}\dot{\Omega} + H_{\text{eff} \, \eta} + \frac{\beta u}{\gamma\delta}\right)
	\end{bmatrix} \, .
\end{eqnarray}

From the LLG equation it is possible to obtain dynamical equations for the angles $\phi$ and $\Omega$ as follows \cite{Mougin-DW}
	
\begin{equation}
    \centering
    \begin{array}{cl}
     \dot{\Omega}=&-\dfrac{\gamma}{M_s}\Gamma_{\Omega} \\
    \\
    \dot{\phi} = &-\dfrac{\gamma}{M_s}\Gamma_{\phi}  \, .
    \end{array}
    \label{eq.mougin}
\end{equation}

Using the DW velocity defined in Eq. \eqref{DW-Vel}, and substituting Eq. \eqref{Torques} in \eqref{eq.mougin}, we obtain 

\begin{equation}
    \centering
    \begin{array}{cl}
     v = &\dfrac{\gamma\delta}{1+\alpha^2}\left(\alpha H_{\text{eff} \, \eta} + \dfrac{u(1+\alpha\beta)}{\gamma\delta} - \dfrac{\Gamma_{\text{eff} \, \eta}}{M_s} \right) \, , \\
     \\
    \dot{\phi} = &\dfrac{\gamma}{1+\alpha^2}\left(H_{\text{eff} \, \eta} +\dfrac{u(\beta-\alpha)}{\gamma\delta}+\dfrac{\alpha}{M_s}\Gamma_{\text{eff} \, \eta}\right)  \, ,
    \end{array}
    \label{eq.dynamics}
\end{equation}

\noindent where $\Gamma_{\text{eff} \, \eta} =  H_{\text{eff} \, \mu}  \cos\phi - H_{\text{eff} \, z}\sin\phi$ is the torque produced by the effective field on the DW center along the $\eta$-direction.

To perform a complete analysis, we need to determine the dipolar and exchange contributions to $\mathbf{H_{\textbf{eff}}}$. The dipolar contribution can be obtained from adopting the shape anisotropy approximation. Under this framework, we can write the dipolar effective field as $\mathbf{H_d} = -4\pi(N_{\mu}M_{\mu}\,\hat{\mu}+N_{\eta}M_{\eta}\,\hat{\eta}+N_{z}M_{z}\,\hat{z})$, where $N_{\mu}$, $N_{\eta}$, and $N_{z}$ are the demagnetizing factors \cite{Aharoni} associated to the $\hat{\mu}$, $\hat{\eta}$ and $\hat{z}$ directions, respectively. The torque produced by $\mathbf{H_d}$ on the DW center is evaluated as $\boldsymbol{\Gamma_d}=-2\pi M_s^{2}(N_{\mu}-N_z)\sin(2\phi)\,\hat{\eta}$. It is worth noticing that $N_{\mu} = N_z$ for NW with circular cross section \cite{Aharoni} and consequently, $\boldsymbol{\Gamma_d}=0$. This result also applies to a NW with a polygonal cross-section with small area  \cite{Altbir-SciRep}. 

Finally, the exchange effective field is given by \cite{Bittencourt-2021} $\mathbf{H_{\text{ex}}} = (2A/M_s)\nabla^2\mathbf{m}$, where $\nabla^2\mathbf{m} = \boldsymbol{\nabla}\left(\boldsymbol{\nabla}\cdot\mathbf{m}\right) - \boldsymbol{\nabla}\times\left(\boldsymbol{\nabla}\times\mathbf{m}\right)$ is the Laplacian operator of $\mathbf{m}$ written in terms of the adopted arbitrary curvilinear system. The explicit equations describing the exchange effective field are presented in Appendix \ref{App-1}. All analytical results developed in this section for a head-to-head N\'eel DW can also be applied to a tail-to-tail DW, just by implementing the transformation $\delta \rightarrow - \delta$.

\subsection{Elliptically bent NW}

From the general model described above we can analyze the DW dynamics in the particular case of an elliptically bent NW whose geometrical description can be done by adopting the following parameterization

\begin{equation}
    \centering
    \begin{array}{cl}
     x = &r \, \text{cosh}\mu \cos \eta \\
     y = &  r \, \text{sinh}\mu \sin \eta \, .
    \end{array}
    \label{parameterization}
\end{equation}

\begin{figure*}[!tbp]
    \centering
    \includegraphics[width=12cm ,angle=0]
    {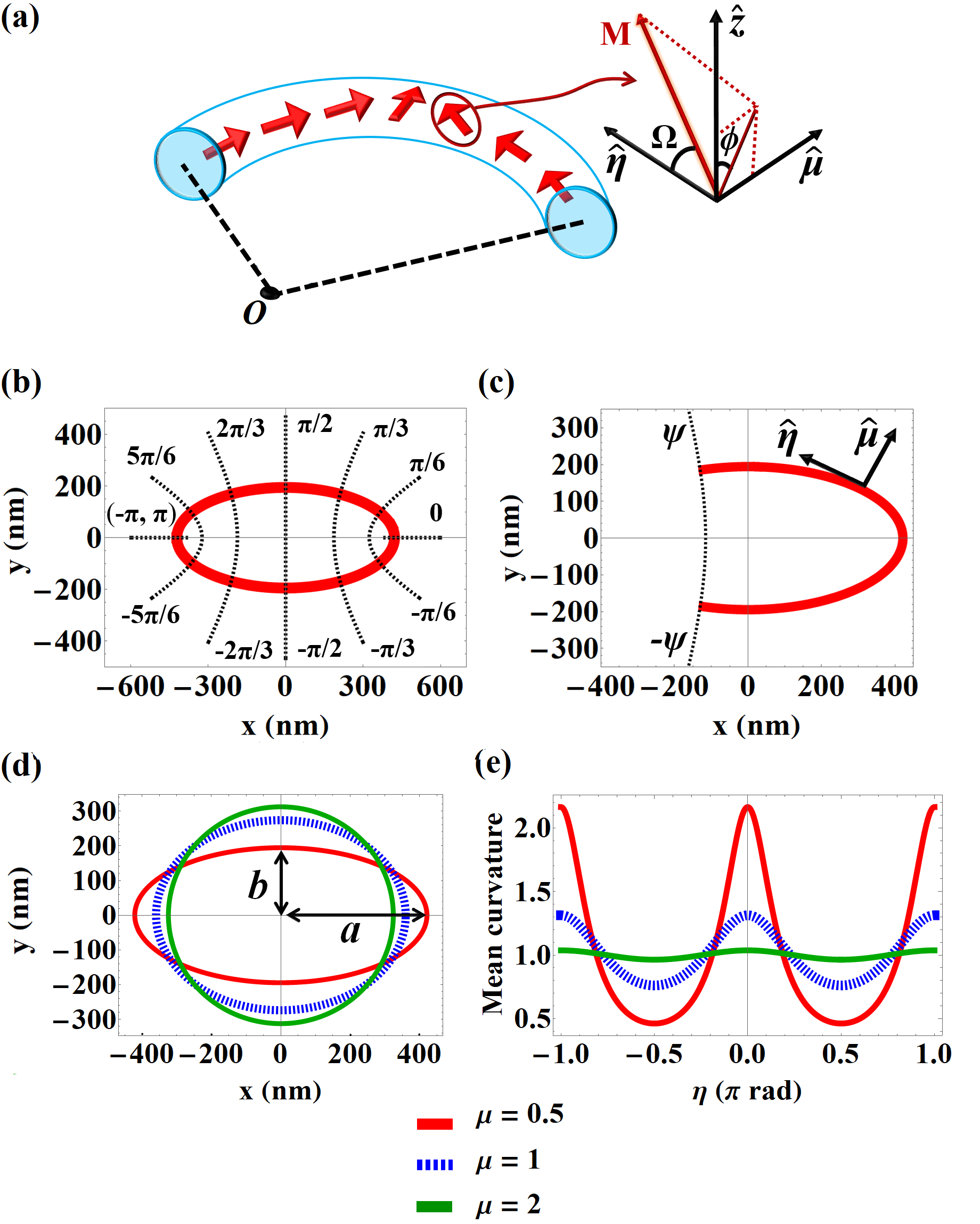}
    \caption{ \textbf{(a)} Representation of the DW magnetization profile lying in an elliptical NW with center located at $O$ (Cartesian origin). 
    \textbf{(b)} Set of hyperbolas (black dashed lines) that intersect orthogonally to the ellipse (red line), defining different values for the $\eta$ coordinate. 
    \textbf{(c)} Representation of the tangential ($\hat{\eta}$) and normal ($\hat{\mu}$) directions on the elliptical NW delimited by the hyperbola segments defined by $\eta = - \psi$ and $\eta = \psi$.
    \textbf{(d)} Illustration of three ellipses with perimeter $\mathcal{P} = 2000$ nm and different  $\mu$ values.
    \textbf{(e)} Curvature $\mathcal{C}$ of complete ellipses as a function  of $\eta$ for different $\mu$ values.}
    \label{fig1}

\end{figure*}

\noindent {Here $a = r\cosh{\mu}$, and $b = r\sinh{\mu}$ are the ellipse semi-axis, with $a \geq b$ and $\mu\geq0$. When the parameter $\mu$ is fixed and $\eta$ varies continuously in the interval $[-\pi, \pi]$, we obtain a complete ellipse. The ellipse eccentricity is intrinsically associated with $\mu$. That is, the greater the value of $\mu$, the smaller the ellipse eccentricity. In this way, the eccentricity vanishes for $\mu \rightarrow \infty$ and the NW geometry presents the shape of a circumference (see Fig. \ref{fig1}-d). On the other hand, when $\eta$ is fixed and $\mu$ varies from $0$ to $\infty$, the parameterization \eqref{parameterization} describes a hiperbola segment that orthogonally intersects the associated ellipse, and the intersection point defines the localization of the $\eta$ coordinate on the ellipse (see Fig. \ref{fig1}-b). In this context, $\eta$ can be considered as an azimuthal-like angle that determines an arbitrary position along the NW length. The unitary vectors that determine the orthogonal ($\hat{\mu}$) and tangential ($\hat{\eta}$) directions are depicted in Fig. \ref{fig1}-c for an arbitrary point on the NW. Explicitly, we have $\hat{\mu} = r\left(\cos{\eta}\sinh{\mu} \, \hat{x} + \sin{\eta}\cosh{\mu} \, \hat{y} \right)/h$ and $\hat{\eta} = r\left(-\sin{\eta}\cosh{\mu} \, \hat{x} + \cos{\eta}\sinh{\mu} \, \hat{y} \right)/h$, where the metric factor of the system is}

\begin{equation}
h \equiv h_{\eta} = h_{\mu} = r\sqrt{\frac{\cosh{(2\mu)} - \cos{(2\eta)}}{2}} \, .
\end{equation}

To properly describe effects due to the geometry on the DW dynamics it is convenient to evaluate the mean curvature $\mathcal{C}$ of the ellipse. Following the formalism presented in Ref. [\onlinecite{Gaididei-PRL}], we obtain $\mathcal{C}(\mu,\,\eta) = \sinh(2\mu)/\left[\cosh(2\mu) - \cos(2\eta)\right]$. The behavior of $\mathcal{C}$ is depicted in Fig.\ref{fig1}-e, where we present the local curvature as a function of $\eta$ for ellipses with fixed perimeter, and $\mu = 0.5$ (red line), $\mu = 1$ (blue-dashed line), and $\mu = 2$ (green-dotted line). It can be noticed that in all cases the maximum curvature is associated with $\eta = 0$ and $\eta = \pi$, while the points presenting minimum local curvature values are located at $\eta = \pi/2$ and $\eta = -\pi/2$. Additionally, due to the decrease in the ellipse eccentricity when $\mu$ increases, $\mathcal{C}$ tends to a constant given by  $\lim_{\mu \rightarrow \infty} \mathcal{C} = 1$. 

It is important to highlight that we are not considering a closed elliptical NW in this work, but a section of an ellipse. If the NW were a complete ellipse, we would have a magnetic NW with two DWs, an issue out of the scope of our work. We analyze specifically bent NWs whose associated ellipses have a fixed perimeter $\mathcal{P} = \int_{-\pi}^{\pi} h\, d\eta = 2000$ nm. Therefore, the NW length $L$ is a fraction of this perimeter that depends on the bounding hyperbola segments characterized by $\eta = \psi$ and $\eta = -\psi$ (see Fig. \ref{fig1}-c), that is, $L = \int_{-\psi}^{\psi} h d\eta$. In this work, we consider $\psi = 0.6\pi$, and then, $\eta$ ranges from $-0.6 \pi$ to $0.6 \pi$. The NW length associated with $\mu = 0.5$, 1.0,  2.0, and  4.0 are respectively $L \approx 1261$ nm, 1224 nm,  1203 nm, and 1200 nm. It is important to notice that the parameter $r$ is intrinsically associated with the ellipse semi-axles $a$ and $b$. Therefore, for a given value of $\mu$ and $\mathcal{P}$, $r$ is implicitly determined.

Now, aiming to perform a subtle analysis on the dynamics of a DW displacing along a bent NW, we determine the exchange effective field $\mathbf{H_{ex}} = (2A/M_s)\nabla^2\mathbf{m}$ acting on the head-to-head Néel DW lying on the elliptic NW. Therefore, from Eqs. \eqref{Ex-F-mu}, \eqref{Ex-F-eta}, and \eqref{Ex-F-z}, we obtain

\begin{eqnarray}
  H_{\text{ex} \, \mu} =  -\frac{2A}{M_s} \Biggr\{2\mathcal{C}^2(\mu,\,\eta_0) \Biggr[\frac{\cos(2\eta_0) + \text{cosh}(2\mu)}{r^2\sinh^2(2\mu)}\sin\phi \nonumber\\ - \frac{\text{sech}\mu}{r\delta}\sqrt{1+\text{csch}^2\mu\sin^2\eta_0}\Biggr]+\frac{\sin\phi}{\delta^2}\Biggr\}\,,\,\,  
\end{eqnarray}

\begin{eqnarray}\label{TangField}
  H_{\text{ex} \, \eta} = -2\sqrt{2}\sin(2\eta_o) \frac{A}{M_sr\delta}\left(\frac{\mathcal{C}(\mu,\,\eta_0)}{\sinh(2\mu)}\right)^{3/2}\, , 
\end{eqnarray}

\noindent and

\begin{eqnarray}
H_{\text{ex} \, z} = -\frac{2A}{M_s\delta^2}\cos\phi \, ,
\end{eqnarray}

\noindent where $\mathcal{C}(\mu,\,\eta_0)$ is the local mean curvature of the ellipse, evaluated at $\eta = \eta_0$ (DW center position) for a given $\mu$. Therefore, one can conclude that in addition to the normal component $H_{\text{ex} \, \mu}$ and the component $H_{\text{ex} \, z}$ of the exchange effective fields (as we observed for DWs on a circular bent NW \cite{Bittencourt-2021}), when a transverse DW is displacing along an elliptical bent NW, it is also under the action of a tangential component of the exchange effective field $H_{\text{ex} \, \eta}$. This tangential field emerges from the exchange-driven curvature-induced effective interactions \cite{Gaididei-PRL}, and vanishes at the positions with maximum and minimum curvatures along the NW length. Because the position with minimum curvature represents the point at which the DW presents its maximum energy (see Fig.\ref{fig2}-(b)), it is an unstable equilibrium point. Therefore, $H_{\text{ex} \, \eta}$ is responsible for inducing a DW motion along the NW even in the absence of external stimuli, bringing it to the stable equilibrium position at the maximum curvature point. This DW pinning has been previously described by Yershov \textit{et al.} \cite{Yershov-2015}. Finally, one can notice that for a circular geometry ($\mu \rightarrow \infty$), which presents a uniform curvature, there is no exchange-driven tangent effective field, that is

$$\lim_{\mu \rightarrow \infty} H_{\text{ex} \, \mu} =  \frac{2A}{M_s}\left[\left(\frac{2}{R\delta}-\left(\frac{1}{\delta^{2}}+\frac{1}{R^{2}}\right)\sin\phi\right)\right]\, ,$$
$$\lim_{\mu \rightarrow \infty} H_{\text{ex} \, \eta} = 0\, , \, \, \, \, \text{and} \, \, \, \, \lim_{\mu \rightarrow \infty} H_{\text{ex} \, z} = -\frac{2A}{M_s\delta^2}\cos\phi \, .$$

\section{DW dynamics in the absence of external stimuli}\label{Result-1}

\begin{figure}
    \centering
    \includegraphics[width=8.5cm ,angle=0]{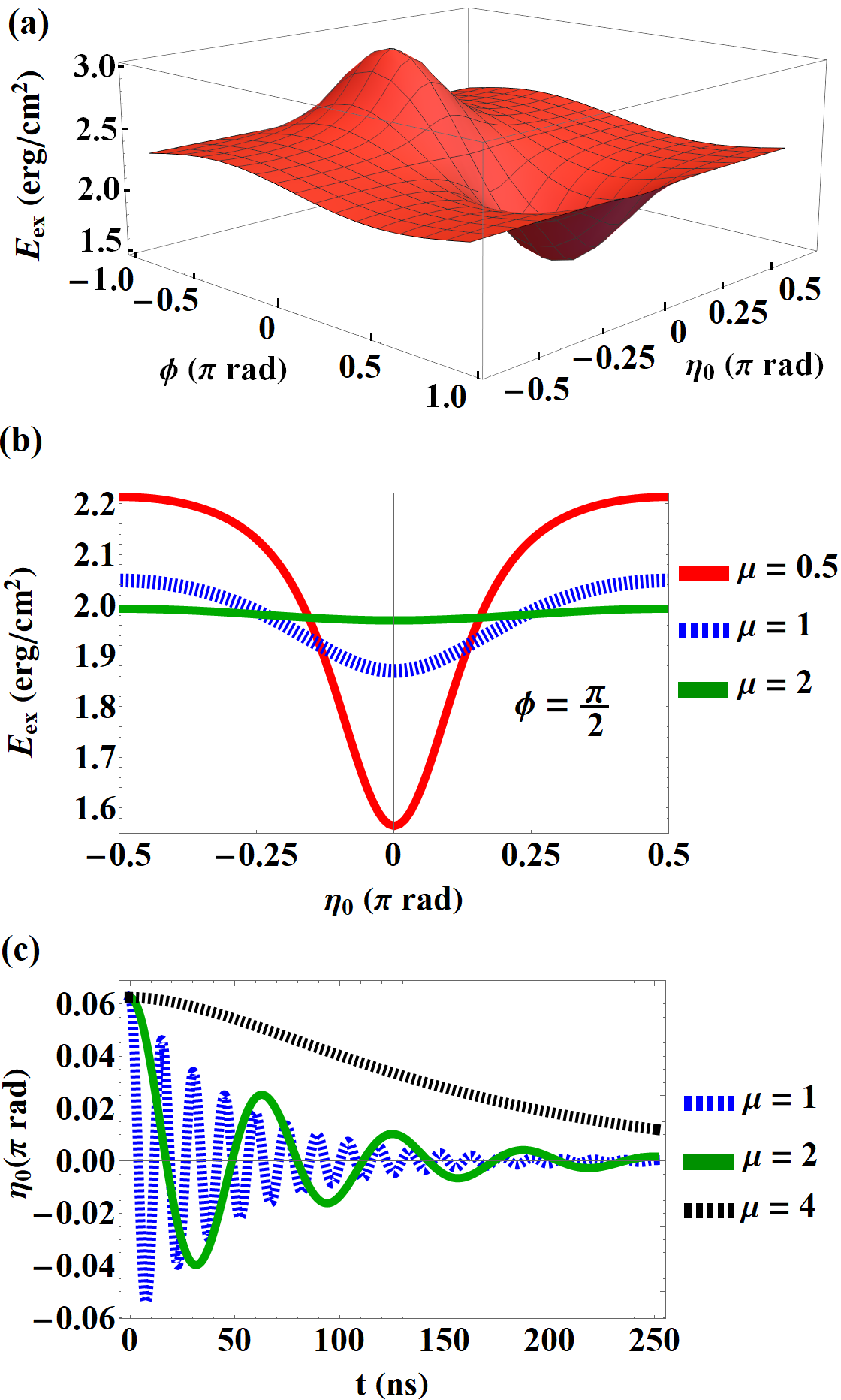}
    \caption{\textbf{(a)} Exchange energy $E_{\text{ex}}$ as a function  of the DW position $\eta_0$ and phase $\phi$ for an elliptical NW with $\mu = 0.5$.
    \textbf{(b)} Behavior of $E_{\text{ex}}$ as a function of the DW position $\eta_0$ for $\phi = \pi/2$.
     \textbf{(c)} Behavior of the DW in the absence of external stimuli developing harmonic decaying oscillations around $\eta_0 = 0$.}
    \label{fig2}
\end{figure}

We are now in position to analyze the DW dynamics on elliptical NWs.  Aiming at having some preliminary perspectives on the DW dynamics, we analyze the DW energy as a function of its position ($\eta_0$) and phase ($\phi$) along the NW. As previously reported, the main contribution to the total energy responsible of the DW dynamic behavior in bent NWs with circular cross-section is the exchange interaction \cite{Moreno,Yershov-2015}. Therefore, to obtain some insights into the  DW behavior, we focus on the exchange energy given by $E_{\text{ex}} = \int_{NW} \mathcal{E}_{ex} dq_{\eta}$, where $\mathcal{E}_{\text{ex}} = -(1/2)\mathbf{M}\cdot\mathbf{H_{ex}}$ is the exchange energy density, and the integral is performed over the NW length. Numerical results for $E_{\text{ex}}$ as a function of $\eta$ are depicted in Fig.\ref{fig2}-a, for $\mu = 0.5$. One can notice that the minimum energy is obtained when the DW is located at $\eta_0 = 0$ (greatest curvature point) and $\phi = \pi/2$ (DW center is pointing outward the bent). Additionally, the maximum energy point corresponds to a DW at $\eta_0 = \pi/2$ (minimum curvature point) and $\phi = -\pi/2$ (DW pointing inward the bent). Therefore, one can conclude that $\eta_0=0$ and $\phi=\pi/2$ represent the equilibrium configuration for the DW, in agreement with results presented in Ref. [\onlinecite{Yershov-2015}]. We have also analyzed the influence of the ellipse eccentricity on the DW exchange energy in the specific case where the DW phase is $\phi = \pi/2$. Main results are depicted in Fig. \ref{fig2}-b for NWs with $\mu = 0.5$ (red line), $\mu = 1.0$ (blue-dashed line), and $\mu = 2.0$ (green line). It can be noticed that the ``potential well'' around $\eta_0 = 0$ and $\phi=\pi/2$ becomes ``deeper'' for more eccentric ellipses, and vanishes for large $\mu$ values.  

Now, we analyze the DW dynamics in the absence of external stimuli ($H = 0$ and $u = 0$). In this case, the DW motion is purely determined by the action of a curvature-induced effective field and the damping effects. Therefore, by considering that the DW is at rest near the maximum curvature point, we  numerically solved  Eqs. \eqref{eq.dynamics} to analyze the DW position and phase as a function of time. Main results are illustrated in Fig. \ref{fig2}-c, from which it is possible to verify that the DW describes harmonic decaying oscillations, reaching its equilibrium position at $\eta_0 = 0$ and $\phi = \pi/2$.

The observed harmonic damped oscillations are associated with the existence of a exchange-driven curvature-induced effective field, which works as a restoring field that tends to bring the system to its equilibrium position. This restoring field can be determined by expanding the tangential component of the exchange field for small displacements around $\eta = 0$, obtaining the so-called harmonic approximation to the tangential exchange field $(H_{\text{ex} \, \eta}^\text{HA}$). Therefore, from Eq. \eqref{TangField}, we obtain

\begin{equation}
H_{\text{ex} \, \eta}^\text{HA} = -\kappa_\mu\eta_0 \, \, \text{,} \, \, \, \, \, \kappa_\mu = \frac{2A}{M_sr\delta}\text{csch}^3\mu \, .
\label{harmonic-field}
\end{equation}

\begin{figure}
    \centering
    \includegraphics[width=8.5cm]{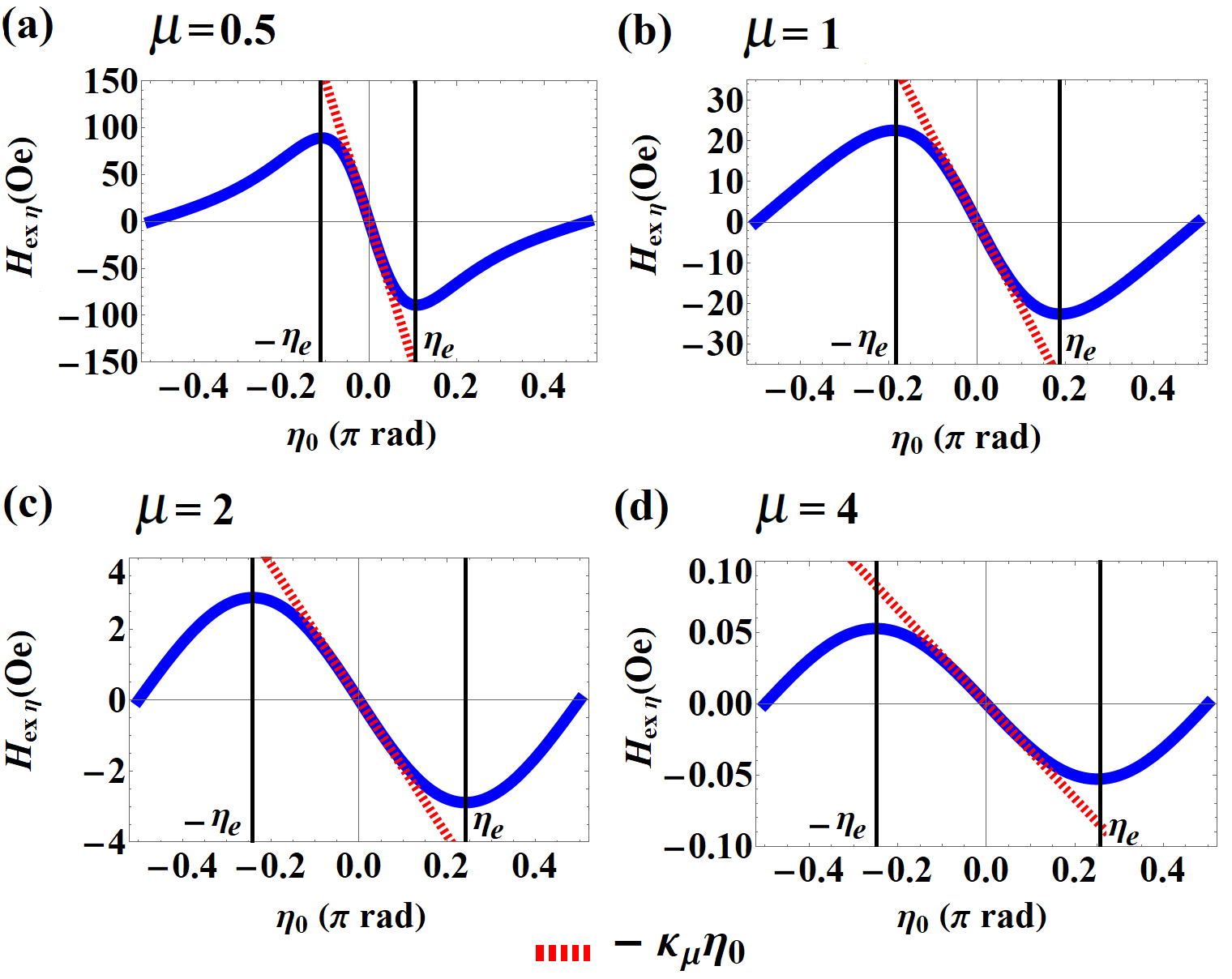}
    \caption{Tangent exchange field $H_{\text{ex} \, \eta}$ as a function  of the DW position $\eta_0$ for $\mu=0.5$ \textbf{(a)}, $\mu=1.0$ \textbf{(b)}, $\mu=2.0$ \textbf{(c)}, and $\mu=4.0$ \textbf{(d)}. The black vertical lines delimit the ``elastic regime'' in which the magnitude of  $H_{\text{ex} \, \eta}$ increases with $\eta_0$. The dashed red line depicts the fitting of the harmonic approximation ($H_{\text{ex} \, \eta}^\text{HA}$) for small DW displacements.}
    \label{fig-exchange-tangent-field}
\end{figure}

Fig. \ref{fig-exchange-tangent-field} illustrates the behavior of $H_{\text{ex} \, \eta}$ (blue line)  
as a function of $\eta_0$ for NWs with different eccentricities. A fit of $H_{\text{ex} \, \eta}^\text{HA}$ (red-dashed line)evidence that the magnitude of $H_{\text{ex} \, \eta}$ increases as a function of $\eta_0$ inside the interval $[-\eta_e, \eta_e]$ (black lines), and the DW moves under the so-called ``elastic regime''. In this case we observe that the harmonic approximation for the restoring field agrees well with $H_{\text{ex} \, \eta}$ even for displacements $\eta_0 \sim 0.1\pi$ rad, mainly for the larger values of $\mu$ ($\mu\geq 1$). Nevertheless, beyond the elastic regime $H_{\text{ex} \, \eta}$ decreases with $\eta_0$ and the harmonic approximation cannot be used to describe the DW motion. One can also notice that the strength of $H_{\text{ex} \, \eta}$ increase as $\mu$ decreases, evidencing the increase of the value of $\kappa_\mu$ as a function of the NW eccentricity. On the other hand, the interval size in which the elastic regime occurs ($\Delta \eta_e \equiv 2\eta_e$) decrease as $\mu$ decreases. 

\begin{figure}
    \centering
    \includegraphics[width=8.5cm]{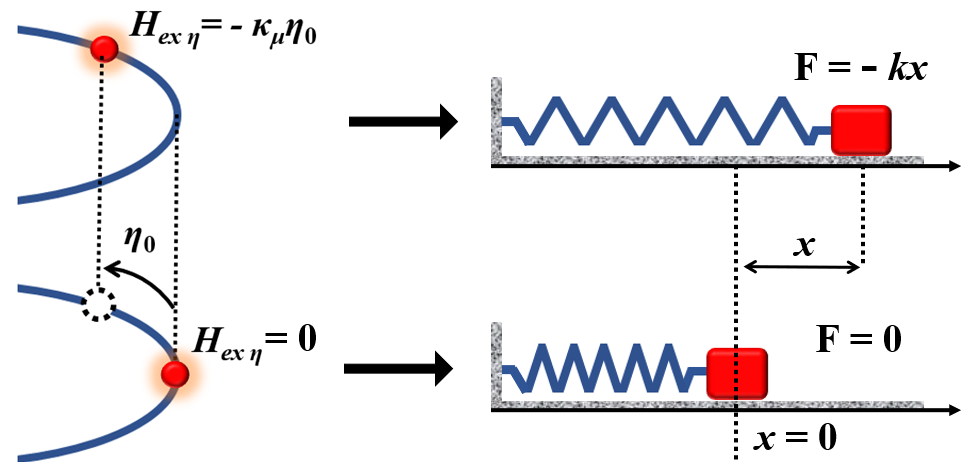}
    \caption{Analogy between the DW motion in a bent NW and a mass-spring system. A DW in the point of maximum curvature corresponds to the spring without deformation. Red dots represent the DW center.}
    \label{M-s-s}
\end{figure}

The above-described results suggested that for small displacements around $\eta = 0$ (maximum curvature point), we can write a dynamic equation for the DW motion as an analogy with a damped harmonic oscillator. That is, under the absence of external stimuli, the DW moves around the point of maximum curvature analogously to a mass-spring system (see Fig. \ref{M-s-s}). If the DW is out of the stable equilibrium point in a region where the linear regime is valid (see Fig. \ref{fig-exchange-tangent-field}), it is under the action of a restoring field, which vanishes in the position of the NW having maximum curvature. In this case, $\kappa_\mu$ can be interpreted an ``elastic constant'', which increases with the NW eccentricity. Therefore, for describing the DW dynamics, we focus on the $\phi$-component of the torque $\Gamma$ given in Eq. \eqref{Torques}, which is  responsible for the fields that have components along the direction tangential to the NW. Because $H = 0$ and $u = 0$, we obtain that $\Gamma_{\phi} = - M_s \left[ (\alpha/\gamma)\dot{\Omega} + H_{\text{ex} \, \eta}\right]$, where $\dot{\Omega} = -(h/\delta)\dot{\eta}_0$. In this case, the only fields acting along the $\eta$-direction are $ H_{\text{ex}}$ and $-(\alpha h/\gamma\delta)\dot{\eta}_0$. Thus, for small displacements around the point of maximum curvature, we can assume that the DW propagates following the Newton's second law. This assumption allows us to write 

\begin{equation}
    m\, \ddot{\eta}_0 = -\kappa_\mu\eta_0 - b_{\mu}\dot{\eta}_0 \, ,
    \label{newtonlaw}
\end{equation}

\noindent where $m$ is the inertial term with dimension of Oe $\times$ s$^2$, and can be interpreted as a DW effective mass. Additionally, $b_{\mu} = (\alpha/\gamma\delta)h \approx (\alpha/\gamma\delta)r\text{ sinh}\mu$. The formal solution of Eq. \eqref{newtonlaw} is 

\begin{equation}
 \eta_0(t) = e^{-t\beta_{\mu}}\left( C_1 e^{t\sqrt{\beta_{\mu}^2-\omega_0^2}} + C_2 e^{-t\sqrt{\beta_{\mu}^2-\omega_0^2}}\right) \, ,
 \label{solution-harmonic}
\end{equation}

\noindent where $\beta_{\mu} = b_{\mu}/2m$ is the damping factor, $\omega_0 = \sqrt{\kappa_{\mu}/m}$ is the natural frequency of the system, that is, the frequency that the oscillator would present in the absence of damping. The constants $C_1$ and $C_2$ can be determined from the initial conditions. For a complete description of the oscillatory character of the DW motion, we need to determine $\omega_0$ from the solutions of Eqs. \eqref{eq.dynamics} in the absence of external stimuli ($H=0$ and $u=0$), and consider $\alpha = 0$. Under these assumptions, for small displacements around $\eta = 0$ and $\phi = \pi/2$, we obtain the DW position as $\eta_0(t) = c_1 e^{i\omega_0 t} + c_2 e^{-i\omega_0 t}$, with $\omega_0$ given by

\begin{equation}
\omega_{0} = \frac{2 A \gamma}{M_s r^2\,\sinh^3\mu} \sqrt{\frac{2r}{\delta}\text{cosh} \mu -\text{coth}^2\mu}\,. 
\label{w0}
\end{equation}

\noindent 
It is worth noticing that $\omega_0$ depends only on intrinsic NW parameters, the magnetic and geometric parameters. As expected, due to the decrease of $\kappa_\mu$ as a function of $\mu$, the natural frequency of the system decreases with the ellipse eccentricity in such a way that the DW oscillatory frequency  practically vanishes for $\mu\approx4$. 

Now, from its classical counterpart, it is also possible to determine the DW effective mass as $m = \kappa_{\mu}/\omega_0^2$, resulting in

\begin{equation}
m = \frac{M_s r^3\text{sinh}^3\mu}{2A\gamma^2\delta\left(\frac{2r}{\delta}\text{cosh} \mu -\text{coth}^2\mu\right)} \, . 
\label{mass}
\end{equation}

\noindent In this case, $m$ increases asymptotically as a function of $\mu$, reaching the maximum value $\lim_{\mu \rightarrow \infty} m = M_s\mathcal{P}^3/\left[16A\pi^2\gamma^2(\mathcal{P}-\pi\delta)\right]\approx6.3\times10^{-16}$ Oe$\times$s$^2$ for $\mu\gtrsim4$, where $\mathcal{P}$ is the perimeter of the ellipse associated, which we kept fixed in our analysis. It is also worth noticing that the obtained expression for the DW mass is not valid for $\mu\lesssim 0.1$. In this case,  $\omega_0$ and $m$ present discontinuities due to the very large eccentricity of the elliptical NW, and the rigid transverse DW assumption is not valid anymore. 

\begin{figure*}
    \centering
    \includegraphics[width=12cm]{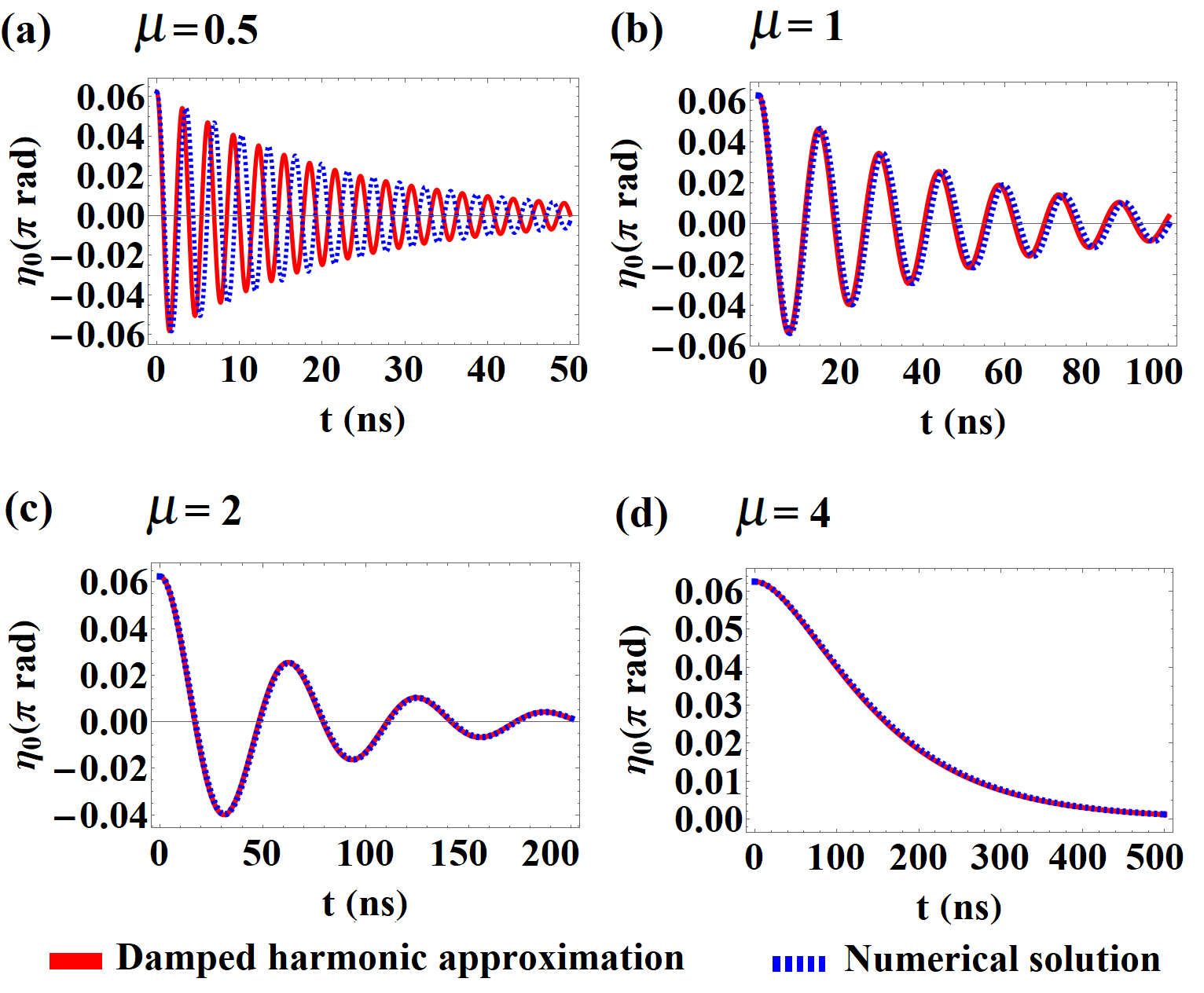}
    \caption{Weak damped oscillations for $\mu = 0.5$ \textbf{(a)}, $\mu=1$ \textbf{(b)}, and $\mu=2$ \textbf{(c)}. The strong damped regime is depicted in \textbf{(d)} for $\mu = 4$. Here, red lines are the solution obtained using the harmonic approximation, while the dashed blue lines are the solutions obtained from the numerical evaluation of the complete dynamics equations. The DW starts from the rest at the position $\eta = \pi/16$.}
    \label{fig-harmonic-sol}
\end{figure*}

Now, to better understand the DW motion described by Eq. \eqref{solution-harmonic} we can perform an analysis for two particular cases regarding the damping. These two cases can be separated by the critical damping   $\beta_\mu=\omega_0$, value at which  the system returns to the steady state as quickly as possible, with no oscillations. Firstly, we investigate the under-damped regime (UD), which occurs for $\beta_{\mu} < \omega_0$, characterized by oscillations with  amplitude gradually decreasing to zero. In this case, the formal solution given by Eq. \eqref{solution-harmonic} can be expressed as damped periodic oscillations described by 

\begin{equation}
\eta_0(t) =  \mathcal{A}e^{-t\beta_{\mu}}\cos\left(t\sqrt{\omega_0^2-\beta_{\mu}^2} - \zeta \right) \, ,
    \label{weakdamping}
\end{equation}

\noindent where $\mathcal{A} = \eta_0(0)\sec\zeta$. Assuming $\dot{\eta}_0(0) = 0$ as the initial condition, we obtain $\zeta = \arccos\left(\sqrt{\omega_0^2-\beta_{\mu}^2}/\omega_0\right) \, .$ In the elastic regime there are positions occupied by the DW where the linear relation between $H_{\text{ex}\,\eta}$ and $\eta$ is not valid. Therefore, we  compared the DW position obtained from the UD harmonic oscillator approach with that predicted by the LLG equation. For this purpose, we  obtained the values of $\eta_0(t)$ from both the analytical approach given by Eq. (\ref{weakdamping}) and the numerical solution of Eqs. \eqref{eq.dynamics} under initial conditions $\eta_0(0) = \pi/16$ and $\phi(0)=\pi/2$. Our results are depicted in Figs. \ref{fig-harmonic-sol}-(a), (b), and (c) for $\mu=0.5$, $1$, and $2$, respectively. One can observe that the comparison between the harmonic oscillator analogy (red line) and the numerical solution of the LLG equation (blue-dashed line) evidences an excellent agreement for $\mu = 1$ and $\mu = 2$. However, for $\mu=0.5$, the frequencies obtained from these two solutions are slightly different in such a way that the predicted DW positions are different. This difference occurs because the length $\Delta\eta_e$ of the elastic regime diminishes as $\mu$ decreases (see Fig. \ref{fig-exchange-tangent-field}). Therefore, a better agreement between the harmonic analogy and the numerical solution of LLG emerges if we consider a smaller initial amplitude. Additionally, our results also evidence the decrease in the DW oscillation frequency as a function of $\mu$. That is, the smaller the ellipse eccentricity, the lower the DW oscillatory frequency. 

The second case we can analyze is the overdamped regime (OD), which is obtained for $\beta_{\mu} > \omega_0$. <in this case, the DW exponentially decays to the equilibrium position with no oscillations. To better understand this OD regime, we need to determine the constants $C_1$ and $C_2$ in Eq. \eqref{solution-harmonic}. Therefore, assuming as  initial condition $\dot{\eta}_0(0) = 0$, we obtain

\begin{eqnarray}
      C_1 = \frac{\eta_0(0)}{2}\left(1+ \frac{\beta_{\mu}}{\sqrt{\beta_{\mu}^2-\omega_0^2}}\right) \, , \\
      C_2 =   \frac{\eta_0(0)}{2}\left(1 - \frac{\beta_{\mu}}{\sqrt{\beta_{\mu}^2-\omega_0^2}}\right) \, .
       \label{strong-damping-sol}
\end{eqnarray}

\noindent We have also compared the results obtained from the damping oscillation approach and the numerical solution of the LLG equation. The results obtained for $\mu = 4$, with initial conditions $\eta_0(0) = \pi/16$ and $\phi(0)=\pi/2$, are depicted in \ref{fig-harmonic-sol}-(d), where one can notice that the DW goes to its equilibrium position with no oscillations. We highlight that larger values of the damping ratio $\beta_\mu$ produce a slow return to the equilibrium position. Therefore, one conclude that the ellipse eccentricity determines the coefficient of the exponential decay of the DW position.
   
It can be observed that the UD and the OD regimes are separated by a critical damping factor $\beta_{\mu} = \omega_0$. Therefore, this relation allows us to find the Gilbert damping of the magnetic NW from the following expression

    \begin{equation}
    \alpha = \frac{4\gamma A \text{csch}^4{\mu_c}}{M_sr^2\omega_0} \, ,
    \label{criticaldamping}
    \end{equation} 

\noindent where $\mu_c$ is associated with the ellipse eccentricity that separates the UD and the OD regimes. Therefore, Eq. \eqref{criticaldamping} provides an interesting way to experimentally determine the Gilbert damping factor of a magnetic NW. That is, by varying the ellipse eccentricity and analyzing the DW oscillations, one can look for the approximated value of $\mu_c$ that produces a transition between the UD and OD regimes. When $\mu_c$ is obtained, Eq. \eqref{criticaldamping} can be used to determine a valid $\alpha$ value.

\section{DW dynamics under a spin polarized electric current}\label{Result-2}

After analyzing the oscillatory behavior of a DW displacing in an elliptic NW in the absence of external stimuli, we can describe the effects that electric currents or magnetic fields generate into the DW dynamics. The extra torque produced on the DW due to the inclusion of an external stimulus can be used for two main purposes: to control the DW equilibrium position, or to unpin the DW and allows it to propagate far from the region of maximum curvature. Therefore, an external stimulus allows a good control of the DW propagation. 

Because most of technological propositions demanding the control of the magnetization properties consider current-driven mechanisms, we will start analyzing the effects of the application of an electric current on the properties of a DW displacing in an elliptical NW. 

\begin{figure*}
\centering
\includegraphics[width=12cm]{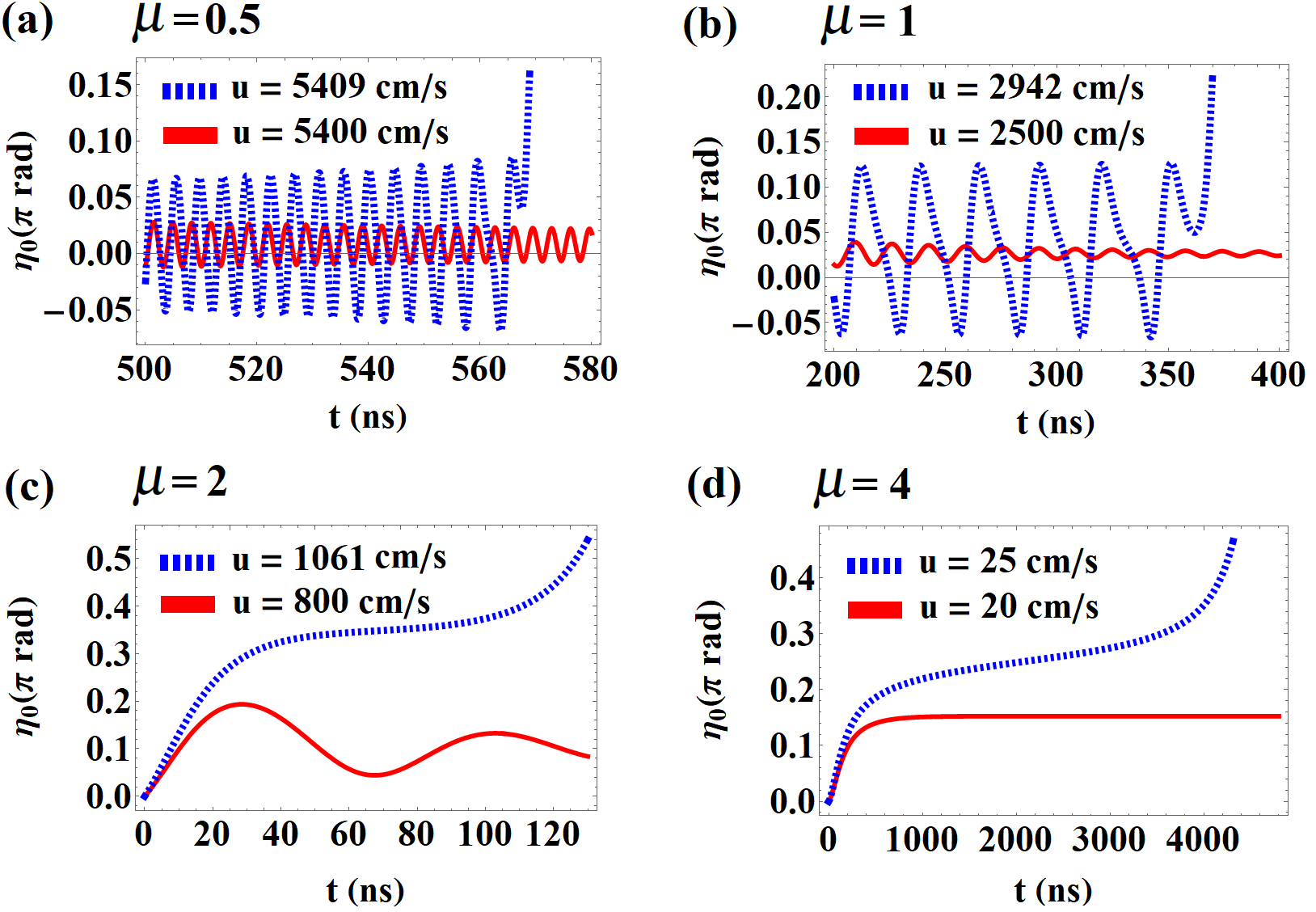}
    \caption{The DW angular position as a function  of time for different   electric current intensities (which corresponds to different $u$ values) for NWs with $\mu = 0.5$ \textbf{(a)}, $\mu = 1$ \textbf{(b)}, $\mu = 2$ \textbf{(c)}, and $\mu = 4$ \textbf{(d)}. Red lines depict the dynamics in the pinning state, while the dashed blue lines refers to the unpinning DW.}
    \label{under-current}
\end{figure*}

The analysis of Eqs. \eqref{eq.dynamics} reveals that the current acts in a different way on the translation and on the rotation motions of the DW. That is, current acts as an effective field $u(1+\alpha\beta)/(\alpha\gamma\delta)$ that influences the DW translation velocity $v$. But the rotation of the DW around the NW axis is affected by a different current-induced effective field, given by $u(\beta-\alpha)/(\gamma\delta)$. This asymmetry in the current effects on the translation and rotation motions of the DW does not allow us to obtain analytical solutions for the DW position and phase. However we can develop a numerical analysis of Eqs. \eqref{eq.dynamics}. Results for the DW position are depicted in Fig. \ref{under-current}, considering four different values of $\mu$ and two values of the electric current. It can be noticed that depending on the current strength, the DW changes its equilibrium position or is unpinned from the region of maximum curvature, moving in an oscillatory motion far from this position. Red lines represent the DW position under the action of a current small enough to keep the DW pinned around the region of maximum curvature. In this case, the DW follows an harmonic damped oscillation around the equilibrium position near the maximum curvature point. In this new equilibrium position, the electric current balances the tangential exchange effective field. The analysis of Fig. \ref{under-current} also reveals that the NW eccentricity influences the frequency, amplitude, and the kind of damped harmonic oscillation (UD or OD). The greater the eccentricity, the greater the oscillation frequency. If the DW is under the action of an electric current that overcomes a threshold value (blue-dashed lines), the DW escapes from the ``potential well'' generated by the exchange-driven curvature-induced effective field. Additionally, we can observe that before being unpinned at the equilibrium position, the DW oscillates around it with a frequency that decreases as a function of $\mu$. Finally, we observe that the amplitude of the oscillations increases with the electric current. Because we have no analytical solutions for describing the DW position as a function of the electric current, the critical values presented in Fig \ref{under-current} were numerically obtained in a trial and error procedure.

\section{DW dynamics under the action of a tangential magnetic field}\label{Result-3}

Due to the asymmetric role played by the spin transfer torque term on the translation and rotation motion, our analysis of the DW motion under electric current is strictly numeric. Therefore, to obtain analytical solutions and better describe the dynamical properties of a DW displacing in an elliptic-shaped NW under the action of external stimulus, we consider now that the system is subject to an external magnetic field $\mathbf{H} = H \hat{\eta}$ tangent to the NW. Although there are experimental difficulties for implementing such a field, our main purpose here is to obtain analytical results that allows us to obtain a description of the DW unpinning phenomenon. 

\begin{figure}
    \centering
    \includegraphics[width=8.5cm]{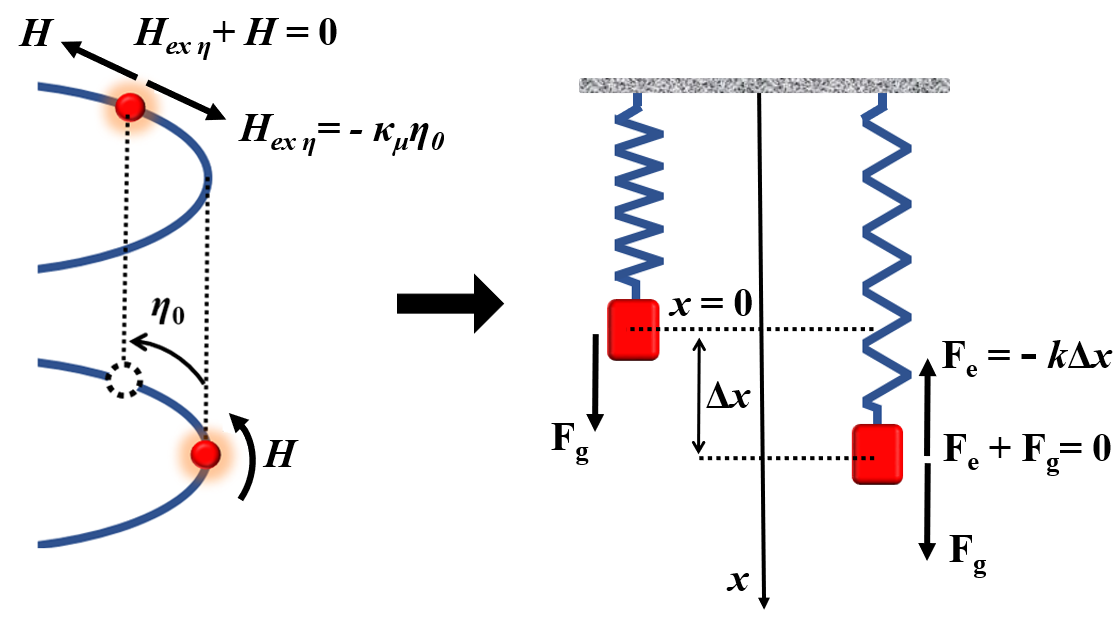}
    \caption{Analogy between the DW motion under the action of a tangential external magnetic field and a mass-spring system under the action of the gravitational force. Red dots represent the DW center. The external magnetic field is responsible for bringing the DW to a new equilibrium point.}
    \label{s-m-g}
\end{figure}

A tangential magnetic field $\mathbf{H}=H\,\hat{\eta}$ applied on the NW can be considered as an extra ``force'' term associated with the external magnetic field in the ``Newton's law'' \eqref{newtonlaw} that describes the DW motion

\begin{equation}
    m\, \ddot{\eta}_0 = H -\kappa_\mu\eta_0 - b_{\mu}\dot{\eta}_0 \, .
    \label{newtonlaw-2}
\end{equation}

\noindent One can notice that the external field plays a role similar to the gravitational force acting on a mass-spring system (see Fig. \ref{s-m-g}). Therefore, the DW assumes a new equilibrium point that depends on the magnetic field strength $H$ and the curvature-induced effective field $H_{\text{ex}\eta}$, which compete with each other. From Eq. \eqref{newtonlaw-2}, we obtain

\begin{eqnarray}
 \eta_{_0H}(t) = e^{-t\beta_{\mu}}\left( \mathcal{K}_1 e^{-t\sqrt{\beta_{\mu}^2-\omega_0^2}} + \mathcal{K}_2 e^{t\sqrt{\beta_{\mu}^2-\omega_0^2}}\right)+\eta_{_H} \, \nonumber\\
 \label{solution-harmonic-under-field}
\end{eqnarray}

\noindent where $\eta_{_H}=\frac{HM_sr\delta\sinh^3\mu}{2A}$. Assuming that the initial conditions are $\eta_0(0) = 0$ and $\dot{\eta}_0(0) = 0$, the constants $\mathcal{K}_1$ and $\mathcal{K}_2$ are evaluated as 

$$ \mathcal{K}_1 = \frac{HM_s\delta\sinh^3\mu}{4A}\left(1 - \frac{\beta_{\mu}}{\sqrt{\beta_{\mu}^2-\omega_0^2}}\right) \, ,$$ 

$$\mathcal{K}_2 = \frac{HM_s\delta\sinh^3\mu}{4A}\left(1 + \frac{\beta_{\mu}}{\sqrt{\beta_{\mu}^2-\omega_0^2}}\right) \, .$$

\begin{figure*}
    \centering
    \includegraphics[width=12cm]{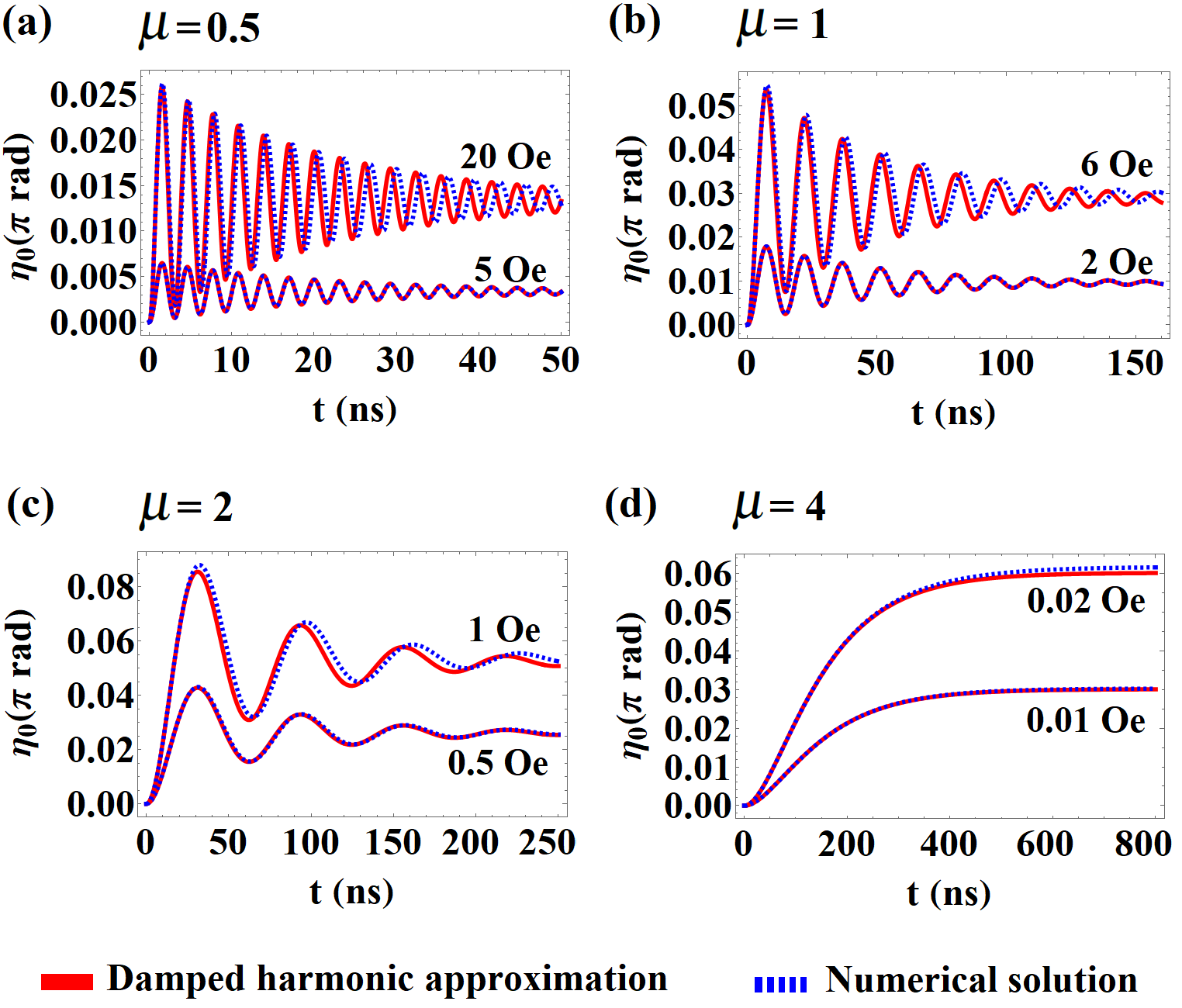}
    \caption{UD regime to the DW oscillations for $\mu = 0.5$ \textbf{(a)}, $\mu=1$ \textbf{(b)}, and $\mu=2$ \textbf{(c)}. The OD regime is depicted in \textbf{(d)} for $\mu = 4$. Here, red lines are the solution described by the harmonic approximation, while the blue-dashed lines are the solutions described by the numerical evaluation of the LLG equation.}
    \label{fig-under-field-harmonic-sol}
\end{figure*}

As expected, the analysis of Eq. \eqref{solution-harmonic-under-field} reveals that the DW presents a damped oscillatory motion around a new equilibrium position $\eta_{_H}$, that depends on the external magnetic field and the ellipse eccentricity. In this context, we can classify again the DW motion under the action of an external magnetic field as UD or OD, for $\beta_\mu<\omega_0$ or $\beta_\mu>\omega_0$, respectively. Previous statements can be corroborated from the analysis of Fig. \ref{fig-under-field-harmonic-sol}, which illustrates the DW position as a function of time obtained from both Eq. \eqref{solution-harmonic-under-field} (red lines) and  numerical solutions og Eq.\eqref{eq.dynamics} (blue-dashed lines). A good agreement between both methods is shown. Again, DWs moving in elliptical NWs with $\mu = 0.5$, $\mu = 1$, and $\mu =2$ present a UD regime, while a OD behavior is obtained for $\mu = 4$. Finally, one can observe that the  magnetic field necessary to bring the DW to a new equilibrium position increases with the ellipse eccentricity. This behavior is intrinsically associated with the increase in  $H_{\text{ex}\eta}$ when $\mu$ decreases.

\begin{figure*}
\centering
\includegraphics[width=12.5cm]{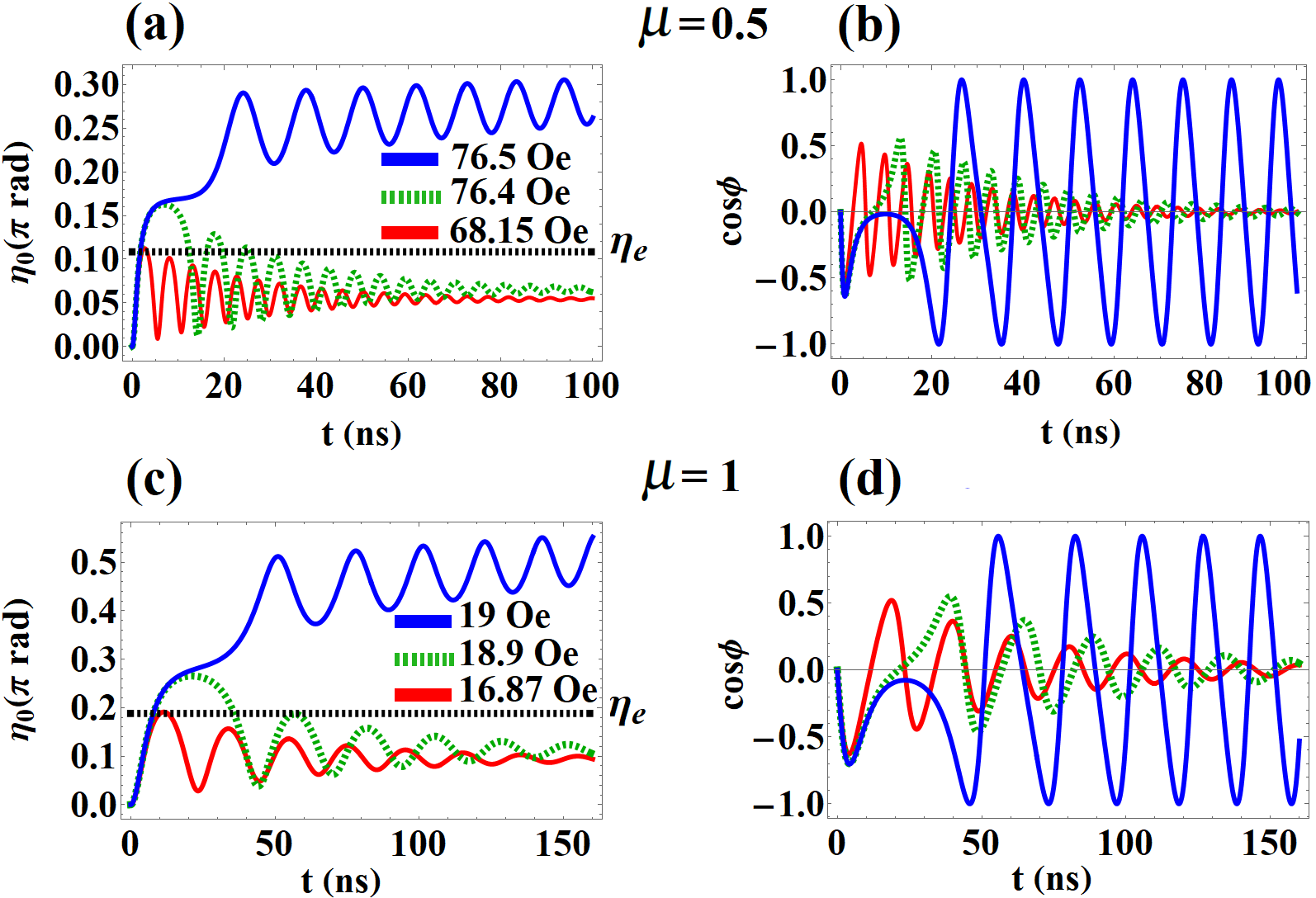}
    \caption{\textbf{(a)} The DW angular position along the time for a NW with $\mu = 0.5$ different intensities of the Zeeman field. 
    \textbf{(b)} The $\cos\phi$ as a function of the time under the same fields as case (a).
  \textbf{(c)} DW angular position along the time for a NW with $\mu = 1$ under different intensities of the Zeeman field.
    \textbf{(d)} The $\cos\phi$ as a function of the time under the same fields as case (c).}
    \label{depinning-field}
\end{figure*}

Because the exchange-driven curvature-induced effective field $H_{\text{ex} \, \eta}$ favours the pinning of the DW around the point of maximum curvature, it is important to highlight the behavior of this field $H_{\text{ex} \, \eta}$ as a function of a position, $\eta$, along the NW. As stated before, $H_{\text{ex} \, \eta}$ increases with the domain wall displacement $\eta_0$ in the interval $[-\eta_e, \eta_e]$, but decreases with $\eta_0$ beyond this region  (see Fig. \ref{fig-exchange-tangent-field}). Therefore, there is a threshold for the external magnetic field from which the DW is unpinned, because $H_{\text{ex} \, \eta}$ diminishes when the DW is in a position out of the elastic regime. In this context, when a small enough magnetic field is applied into the system, the DW  follows harmonic decaying oscillations with its equilibrium position occurring when $H + H_{\text{ex} \, \eta} = 0$. Assuming that when the external magnetic field is applied into the system the DW is pinned in the region of maximum curvature, it starts oscillating around the new equilibrium position. Under this assumption, the limit field ensuring that the DW does not exceed the elastic regime can be estimated as $H_e \approx -H_{\text{ex} \, \eta}( \eta_e/2)$, in such a way that the DW equilibrium position is $\eta_0 = \eta_e/2$. In this case, in the first oscillation, the DW displaces  $\Delta\eta_0 \approx \eta_e/2$ beyond the equilibrium position and then, the DW position does not overcome the elastic limit. After some algebra, $H_e$ can be evaluated as

\begin{equation}\label{Dep-Field}
    H_e \approx \dfrac{4A\sqrt{2+2\text{cosh}(2\mu)-\lambda}}{M_sr\delta\left(2\text{cosh}(2\mu)-\sqrt{2-2\text{cosh}(2\mu)+\lambda}\right)^{3/2}} \, ,
\end{equation}

\noindent where $\lambda = \sqrt{14+2\text{cosh}(4\mu)}$. Eq. \eqref{Dep-Field} allows us to estimate the unpinning field, $H_e$, as a function of the NW eccentricity. For instance, $H_e \approx 68.15$ and $16.87$ Oe for $\mu = 0.5$ and $1$, respectively. To analyze the DW behaviour for different external magnetic fields and NW eccentricities, we determine the DW position and phase for elliptically-shaped NW with eccentricities $\mu=0.5$ and $\mu=1$. The obtained results are depicted in Fig. \ref{depinning-field}, where red lines represent the DW position and phase when it is under the action of the unpinning magnetic field, $H_e$. In this case, the DW reaches a new equilibrium position depending on the NW eccentricity, given by $\eta_e/2\approx 0.05$ and $0.1$ for $\mu=0.5$ and $1$, respectively. We also highlight that during its first oscillation, the DW does not overcome the position $\eta_e$, delimiting the elastic regime (see black-dashed lines in Figs. \ref{depinning-field}). The green-dotted lines in Figs. \ref{depinning-field}-(a) and (c) show the DW position when it is under the action of a magnetic field whose strength is slightly higher than $H_e$. In this case, during its oscillations, the DW reaches a position along the NW that is out of the range in which the elastic regime occurs. Nevertheless, it is noteworthy that even if the DW exceeds the position $\eta_e$ during the first oscillation, the curvature-induced tangential effective field is still big enough to ensure that $|H_{\text{ex} \, \eta}| > H$, which brings the DW back to a position where the elastic regime is valid. Finally, if the magnetic field overcomes a threshold, the DW is unpinned and starts propagating along the NW under the Walker regime. That is, its position increases as a function of time in an oscillatory motion followed by a DW  rotation. This critical magnetic field was numerically evaluated, the DW position and phase are depicted by blue lines in  Fig.\ref{depinning-field}. 

We also highlight that the analysis of the DW phase yields a completely analogous behavior to that for the DW position (see Figs.\ref{depinning-field}-(b) and (d)). In this case, While the DW position in the damped oscillatory motion behaves like a mass-spring system, we observe that the oscillations in the DW phase behave as a torsion pendulum that performs a damped harmonic oscillation around the equilibrium phase. Furthermore, the natural frequency $\omega_0$ and the damping parameter $\beta_{\mu}$ are the same as obtained for the DW translational motion.

\section{Conclusions}\label{Conclusions}

In this work we develop an analysis of a transverse domain wall dynamics displacing along  a nanowire with circular cross-section and bent in an elliptical shape. The curvature gradient of the elliptical geometry strongly affects the DW translation and rotation motions. Particularly, the exchange-driven curvature-induced tangential effective field produces a pinning potential, which increases with the ellipse eccentricity and traps the DW nearby the region presenting maximum curvature. The DW equilibrium position is a result of the competition between the tangential effective field and the torque on the magnetization originated from an external stimulus (electric current of magnetic field). Our results evidence that if the external stimuli are bellow a threshold value, the DW moves following a damped harmonic oscillation around the equilibrium position. Nevertheless, if the external stimulus overcomes the threshold imposed by the exchange-driven curvature-induced tangential effective field, the DW is unpinned from the region with maximum curvature and moves along the NW with an oscillatory motion characteristic of the Walker regime. Therefore, the results here presented support the possibility of controlling the DW dynamic  by varying both the NW curvature and external stimulus.

\begin{widetext}
\appendix

\section{Expressions for the exchange energy density and exchange field in an arbitrary curvilinear NW}\label{App-1}  

The exchange energy can be calculate as  $E_{\text{ex}} = \int_{NW} \mathcal{E}_{\text{ex}} h_{\eta} d\eta$, where $\mathcal{E}_{\text{ex}} =-(1/2) \mathbf{M}\cdot\mathbf{H_{ex}}$ is the exchange energy density and $\mathbf{H_{\text{ex}}}=(2A/M_s)\nabla^2\mathbf{m}$ is the exchange effective field evaluated in an arbitrary point on the NW length. Under this framework, we assume that the magnetization varies only along to the tangential direction, that is, $\mathbf{m} = \mathbf{m}(\eta)$ and then the DW is described by $\Omega = \Omega(\eta)$. Furthermore, we recall that the considered NW lies in the $xy$-plane, which implies $h_z = 1$. Thus, we can determine an expression for the exchange density energy in a transverse DW lying in a general curvilinear system ($\mu$, $\eta$, $z$) as follows

\begin{eqnarray}
\mathcal{E}_{\text{ex}} = A \Biggr\{ \left(\frac{1}{{h_{\eta}}}\frac{\partial \Omega}{\partial \eta}\right)^2 - \frac{2}{h_{\mu}h_{\eta}^2}\frac{\partial h_{\eta}}{\partial \mu}\frac{\partial \Omega}{\partial \eta}\sin\phi +
\frac{3 - \cos(2\phi) + 2\cos(2\Omega)\cos^2\phi}{4h_{\mu}^2h_{\eta}^2}\biggr[\left(\frac{\partial h_{\mu}}{\partial \eta}\right)^2 + \left(\frac{\partial h_{\eta}}{\partial \mu} \right)^2 -h_{\eta}\frac{\partial^2 h_\eta}{\partial \mu^2} \biggr]  \nonumber\\
  +  \frac{3 + \cos(2\Omega) - 2\sin^2\Omega\cos(2\phi)}{4 h_{\mu}h_{\eta}^3} \left(\frac{\partial h_{\mu}}{\partial \eta}\frac{\partial h_{\eta}}{\partial \eta} - h_{\eta}\frac{\partial^2 h_{\mu}}{\partial \eta^2} \right) 
    + \frac{1}{h_{\mu}^3h_{\eta}}\frac{\partial h_{\mu}}{\partial \mu}\frac{\partial h_{\eta}}{\partial \mu} \left( \cos^2\Omega + \sin^2\Omega\sin^2\phi \right) \Biggr\} .
  \label{Energy-density-general}
\end{eqnarray}

Therefore, it is useful to determine an expression for the exchange effective field acting on a DW lying in a bent NW. For this purpose, we need to calculate the Laplacian operator of $\mathbf{m}$ and evaluate it at the DW center. Specifically, for a head-to-head DW, we have

\begin{eqnarray}\label{Ex-F-mu}
  H_{\text{ex} \, \mu} = -\frac{2A}{M_s}\Biggr\{\Biggr[\frac{1}{\delta^2} + \left(\frac{1}{h_{\mu}h_{\eta}}\frac{\partial h_{\mu}}{\partial \eta}\right)^2 + \left(\frac{1}{h_{\mu}h_{\eta}}\frac{\partial h_{\eta}}{\partial \mu}\right)^2 + \frac{1}{h_{\mu}h_{\eta}^3} \frac{\partial h_{\mu}}{\partial \eta}\frac{\partial h_{\eta}}{\partial \eta} + \frac{1}{h_{\mu}^3h_{\eta}}\frac{\partial h_{\mu}}{\partial \mu}\frac{\partial h_{\eta}}{\partial \mu}  - \frac{1}{h_{\mu}h_{\eta}^2}\frac{\partial^2 h_{\mu}}{\partial \eta^2} \nonumber\\  - \frac{1}{h_{\mu}^2h_{\eta}}\frac{\partial^2 h_{\eta}}{\partial \mu^2}\Biggr]\sin\phi - \frac{2}{\delta h_{\mu}h_{\eta}}\frac{\partial h_{\eta}}{\partial \mu} \Biggr\} \, ,
\end{eqnarray}


\begin{eqnarray}\label{Ex-F-eta}
  H_{\text{ex} \, \eta} = -\frac{2A}{M_s}\Biggr\{\Biggr[\frac{1}{h_{\mu}h_{\eta}^3}\frac{\partial h_{\eta}}{\partial \eta}\frac{\partial h_{\eta}}{\partial \mu}-\frac{1}{h_{\mu}^3h_{\eta}}\frac{\partial h_{\mu}}{\partial \eta}\frac{\partial h_{\mu}}{\partial \mu} +  \frac{1}{h_{\mu}^2h_{\eta}}\frac{\partial}{\partial \mu}\left(\frac{\partial h_{\mu}}{\partial \eta}\right) \nonumber \\- \frac{1}{h_{\mu}h_{\eta}^2}\frac{\partial}{\partial \mu}\left(\frac{\partial h_{\eta}}{\partial \eta}\right) \Biggr]\sin\phi + \frac{1}{\delta h_{\mu}h_{\eta}}\frac{\partial h_{\mu}}{\partial \eta} \Biggr\} \, ,
\end{eqnarray}


\noindent and

\begin{eqnarray}\label{Ex-F-z}
  H_{\text{ex} \, z} = -\frac{2A}{M_s\delta^2}\cos\phi \, .
\end{eqnarray}

It is worth noticing that if the metric factors $h_{\eta}$ and $h_{\mu}$ are $\eta$-independent, $\partial h_{\eta}/\partial\eta = \partial h_{\mu}/\partial\eta = 0$, and then $H_{ex \, \eta} = 0$. Therefore, we can state that the tangential exchange field is directly associated with metric variations along the length of the NW. In other words, $H_{ex \, \eta}$ is induced by the curvature gradient along the $\eta$-direction. Furthermore, we note that $h \equiv h_{\eta} = h_{\mu}$ (which occurs in the case of the elliptical-shaped NW, for example), the tangential component of the exchange-driven curvature-induced effective field becomes independent of the DW phase $\phi$, that is 

\begin{eqnarray}\label{Ex-F-eta-particular-case}
  H_{\text{ex} \, \eta} = -\frac{2A}{M_s\delta h^2}\frac{\partial h}{\partial \eta} \, .
\end{eqnarray}

We highlight that our results also applie to a tail-to-tail DW by performing the change $\delta \rightarrow - \delta$.

\end{widetext}

\end{document}